\documentclass[10pt,twoside]{hsqcd}
\usepackage{epsf,amsmath}
\usepackage{graphicx}

\newcommand{\bean}{\begin{eqnarray*}}
\newcommand{\eean}{\end{eqnarray*}}
\newcommand{\gapproxeq}{\lower
.7ex\hbox{$\;\stackrel{\textstyle >}{\sim}\;$}}
\newcommand{\lapproxeq}{\lower
.7ex\hbox{$\;\stackrel{\textstyle <}{\sim}\;$}}

\newcommand\lsim{\mathrel{\rlap{\lower4pt\hbox{\hskip1pt$\sim$}}
    \raise1pt\hbox{$<$}}}
\newcommand\gsim{\mathrel{\rlap{\lower4pt\hbox{\hskip1pt$\sim$}}
    \raise1pt\hbox{$>$}}}
\newcommand{\ba}{\begin{array}}
\newcommand{\ea}{\end{array}}

\newcommand{\be}{\begin{equation}}
\newcommand{\ee}{\end{equation}}
\newcommand{\bear}{\begin{eqnarray}}
\newcommand{\eear}{\end{eqnarray}}
\newcommand{\tab}{\hspace*{0.5cm}}

\newcommand{\ket}{\,\rangle}
\newcommand{\bra}{\langle \,}
\newcommand{\eqn}[1]{(\ref{#1})}
\newcommand{\cO}{{\cal O}}
\newcommand{\bel}[1]{\be\label{#1}}

\newcommand{\mB}{\mathcal{B}}

\newcommand{\Frac}[2]{\frac{\displaystyle #1}{\displaystyle #2}}
\newcommand{\Int}{\displaystyle{\int}}

\begin{document}

\title{$V+A$ and $V-A$ Correlators at Large $N_C$: \\
From OPE to Resonance Theory 
\thanks{
I would like to thank the organisers of HSQCD'05 (St. Petersburg, Russia) 
for their hospitality.    
This work was  supported by    
EU~RTN~Contract~CT2002-0311. 
This talk  is based on
Ref.~\cite{correlator}
}}

\author{\underline{J.~J.~Sanz-Cillero}\\ \\
Groupe Physique Th\'eorique, IPN Orsay \\
Universit\'e Paris-Sud XI,  91406 Orsay, France \\
E-mail:  cillero@ipno.in2p3.fr }

\maketitle


%
%
%
%
\begin{abstract}
\noindent 
The spin--1  correlators are analysed  in this talk through  
a large $N_C$ resonance theory. 
The matching to perturbative QCD and the first terms in the OPE  
constrains  the hadronic parameters. A further sum-rule 
analysis  shows the wider range of validity of the resonance description, 
which can help to discern the proper structure of the QCD mass spectrum.  
\end{abstract}

\markboth{\large \sl \underline{J.~J.~Sanz-Cillero} 
\hspace*{2cm} HSQCD 2005} {\large \sl \hspace*{1cm} }

\section{Introduction}
The purely perturbative Quantum Chromodynamics
calculations (pQCD) and its Operator Product Expansion (OPE) 
are essential tools to describe   the strong
interactions~\cite{Shifman}.  
However, they stop being valid 
as  we enter into the non-perturbative QCD regime 
and  one needs to consider  alternative
descriptions keeping, nevertheless, the agreement 
to  OPE at high energies.  A resonance theory with infinite  
narrow--width resonances has been shown to 
fulfill these requirements in the large $N_C$ limit~\cite{NC}, being $N_C$ the
number of colours. 
Within this framework, we will consider a matching to pQCD and OPE up to
$\cO(\alpha_s)$.

In this talk, we analyse the two--point Green-functions,  
\be
(q^\mu q^\nu - q^2 g^{\mu\nu})\, \Pi(q^2)_{_{XY}} \quad 
= \quad i \,\,  \Int\,\, dx^4 \,\, e^{i \, q\,  x} \,\,\, 
\bra T\{J_{_X}^\mu(x)\,J_{_{Y}}^\nu(0)^\dagger \,\} \ket \, , 
\ee
with $J_X^\mu$ and $J_Y^\nu$ denoting either a vector or an axial-vector
current. We actually analyse the $V-A$ and $V+A$ combinations,
respectively $\Pi_{_{LR}}=\Pi_{_{VV}}-\Pi_{_{AA}}$ and
$\Pi_{_{LL}}=\Pi_{_{VV}}+\Pi_{_{AA}}$. Only  the light quarks
$u/d/s$ are considered and within the chiral and large $N_C$ limits.

Several structures $\{M_n^2\}$ for the 
resonance mass spectrum are explored  in Ref.~\cite{correlator}.
Here we  refer just to the Regge-like models and the
spectrum of the 5D--holographic models.

\section{Fixing the resonance parameter through pQCD and OPE} 
In the deep euclidean region 
$Q^2=-q^2 \gg \Lambda_{_{QCD}}^2$,  the correlators 
are given by  the OPE power series~\cite{Shifman}, 
\bel{eq.OPEcorrelator}
\Pi_{_{LL}}^{^{OPE}}\,= \, \Pi_{_{LL}}^{^{pQCD}}
\, + \, \displaystyle{\sum_{m=2}^\infty} \Frac{\bra
\cO_{_{(2m)}}^{^{LL}}\ket}{Q^{2m}} \, , 
\qquad \qquad
\Pi_{_{LR}}^{^{OPE}}\,= \, 
\displaystyle{\sum_{m=3}^\infty} \Frac{
\bra \cO_{_{(2m)}}^{^{LR}}\ket}{Q^{2m}} \, .
\ee
The $\cO(\alpha_s^2)$ corrections are
dropped in this work so the $1/Q^{2m}$ coefficients 
are just the constant 
condensates $\bra \cO_{_{(2m)}}\ket$.

On the other hand, the correlators 
are provided  at large $N_C$ by the infinite series,   
\bear
\Pi_{_{LL}}^{^{N_C\to\infty}}\,= \, \, \Frac{F_\pi^2}{Q^2} \, + \,
\displaystyle{\sum_{j=1}^\infty} \Frac{F_j^2}{M_j^2+Q^2} \, , 
\qquad 
\Pi_{_{LR}}^{^{N_C\to\infty}}\,= \, - \, \Frac{F_\pi^2}{Q^2} \, + \, 
\displaystyle{\sum_{j=1}^\infty} (-1)^{j+1}\,\,\Frac{F_j^2 }{M_j^2+Q^2} \, , 
\eear
where a resonance 
spectrum with alternating parity is assumed, being the first one, $j=1$,   
the vector $\rho(770)$.
The masses are ordered on increasing order ($M_1\leq M_2\leq...$).

\subsection{Step 1: matching pQCD}
\tab 
In order to recover the leading contribution in
$\Pi_{_{LL}}$, provided by pQCD at $Q^2\gg \Lambda_{_{QCD}}^2$, 
one needs to impose  the constraint~\cite{correlator,EspriuRegge}:
\bel{eq.relationpQCD}
F_j^2\quad =\quad \delta M_j^2 \,\cdot \, 
\left[ \, \Frac{1}{\pi}\mbox{Im}\Pi(M_j^2)_{_{LL}}^{^{pQCD}}
\quad + \quad \cO\left(\Frac{\Lambda_{_{QCD}}^2}{M_j^2}\right)
\, \right]\, ,  
\ee
with 
$\delta M_j^2\equiv M_{j+1}^2-M_{j}^2$. 
The logarithmic behaviour from  
$\Frac{1}{\pi}\mbox{Im}\Pi(M_j^2)_{_{LL}}^{^{pQCD}}$ ensures 
the dominant log dependence 
$\Pi_{_{LL}}^{^{N_C\to\infty}}\simeq \Pi_{_{LL}}^{^{pQCD}}$ 
at high $Q^2$. The   
$\cO\left(\frac{\Lambda_{_{QCD}}^2}{M_j^2}\right)$  
corrections are suppressed for large masses and they are here neglected for
$M_j^2\gsim 2$~GeV$^2$. 

This matching relation breaks down    
for the resonances laying by the
non-perturbative regime of QCD. The lightest states,  
$\pi, \, \rho(770)$ and $a_1(1260)$,  
need to be considered apart into  ``\,non-perturbative''\,
sub-series 
\be
\Delta \Pi_{_{LL}}^{^{non-p.}}= \Frac{F_\pi^2}{Q^2}  + 
 \Frac{F_\rho^2}{M_\rho^2+Q^2} +  
\Frac{F_{a_1}^2}{M_{a_1}^2+Q^2}  , 
\quad
\Delta \Pi_{_{LR}}^{^{non-p.}}=  -  \Frac{F_\pi^2}{Q^2}  + 
 \Frac{F_\rho^2}{M_\rho^2+Q^2} -  
\Frac{F_{a_1}^2}{M_{a_1}^2+Q^2} , 
\label{eq.MHA}
\ee
The couplings $F_\rho, \, F_{a_1}$ and the mass $M_{a_1}$ are
left  as free  parameters, whereas  
$F_\pi=92.4$~MeV and  $M_\rho=0.77$~GeV are taken  as inputs.

An asymptotic structure is assumed for the masses  
with $M_n^2\gsim 2$~GeV$^2$ ($n\geq 3$).    
In this talk we refer just to the Regge spectrum~\cite{EspriuRegge,LVS,toymodel}, 
with equal squared mass interspacing, 
$M_{n}^2 = \Lambda^2 + n\, \delta M^2$,  
and the 5D--spectrum~\cite{Son,5D}, 
with equal mass interspacing, $M_{n} = \Lambda+ n\, \delta M$.
The parameters are set such that $M_3=M_{\rho'}\simeq 1.45$~GeV and 
$M_4=M_{a_1}\simeq 1.64$~GeV~\cite{PDG}.  
This fixes the couplings  $\{F_n^2\}_{n\geq 3}$  through
Eq.~\eqn{eq.relationpQCD} and, hence, the ``\,perturbative''\, contribution  
$\Delta \Pi^{^{pert.}} \, = \, \Pi^{^{N_C\to\infty}}\, - \, 
\Delta \Pi^{^{non-p.}}$.

\subsection{Step 2: matching the leading OPE power behaviour}
\tab 
The second step is  to match power behaviour of the 
first non-trivial operator in the OPE, 
this is,  to demand 
\bel{eq.relationOPE}
\bra \cO_{_{(2)}}^{^{LL}}\ket = 
\bra \cO_{_{(2)}}^{^{LR}}\ket = \bra \cO_{_{(4)}}^{^{LR}}\ket = 0
\, , \qquad \mbox{with} \quad  
\bra \cO_{_{(2m)}}\ket= \Delta \bra \cO_{_{(2m)}}\ket^{^{non-p.}}+ 
\Delta \bra \cO_{_{(2m)}}\ket^{^{pert.}}
\, . 
\ee
The contributions  $\Delta \bra \cO_{_{(2m)}}\ket^{^{non-p.}}$  
from $\Delta \Pi^{^{non-p.}}$   are obtained by  
expanding Eq.~\eqn{eq.MHA} in powers of   $M_j^2/Q^2$. 
The contributions  $\Delta \bra \cO_{_{(2m)}}\ket^{^{pert.}}$  
come from the ``\,perturbative''\, sub-series $\Delta \Pi^{^{pert.}}$.   
Taking  $\Pi^{^{pQCD}}$  and the asymptotic spectrum 
$\{M_n^2\}_{n\geq 3}$ as inputs, 
one gets $\{F_n^2\}_{n\geq 3}$  through Eq.~\eqn{eq.relationpQCD} 
and finds that the combination 
$\left(\Delta \Pi^{^{pert.}}-\Pi^{^{pQCD.}}\right)$ shows the power structure 
$\displaystyle{\sum_{m=1}^{\infty} }
\Frac{\Delta \bra \cO_{_{(2m)}}\ket^{^{pert.}}}{Q^{2m}}$.  
The trivial expansion in powers of $M_j^2/Q^2$ is not valid here  
since there is always an infinite number of 
states with $M_j^2> Q^2$.  
The first coefficients  $\Delta \bra \cO_{_{(2m)}}\ket^{^{pert.}}$ 
are recovered through a numerical analysis in the range
$Q^2=2\, $--$\, 6$~GeV$^2$, together with a consistent estimate 
of their uncertainties~\cite{correlator}.

By   imposing the OPE constraints in Eq.~\eqn{eq.relationOPE},   
$F_\rho, \, M_{a_1}$ and $F_{a_1}$  become fixed 
and, hence, the correlators $\Pi^{^{N_C\to\infty}}$ result fully determined.

\begin{figure}[h!]
\begin{center}
\includegraphics[width=6cm,clip]{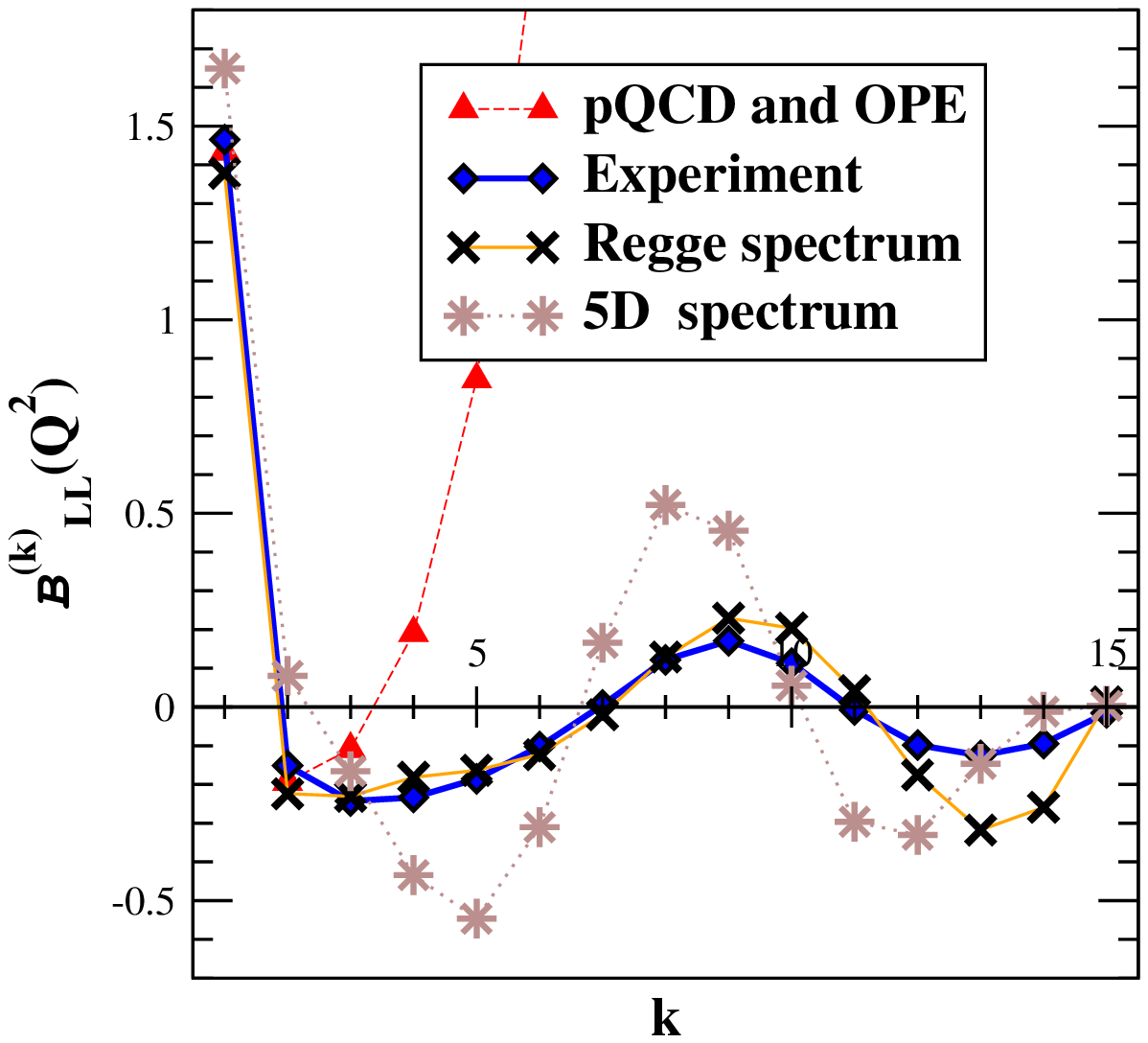}
\hspace*{0.5cm}
\includegraphics[width=5.5cm,clip]{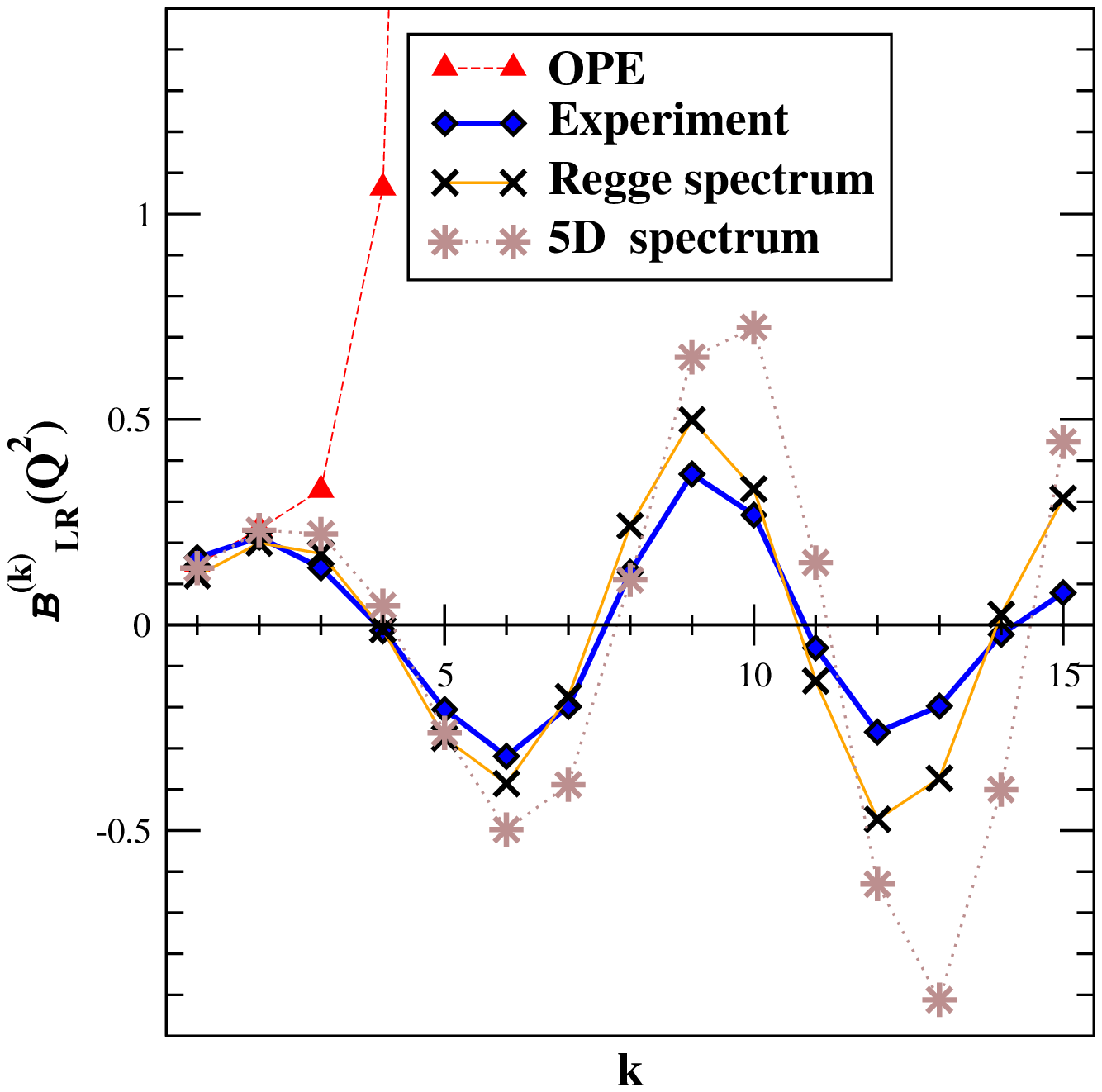}
\caption{Moments $\mB^{^{(k)}}(Q^2)$  for $Q^2=3$~GeV$^2$. The 
experimental determination is compared to that from  the  
large $N_C$ resonance theories  and OPE~\cite{correlator}.}
\label{fig.B}
\end{center}
\end{figure}

\section{Discerning between resonance models: sum--rule analysis}
\tab
The matching to pQCD and the leading OPE operators 
can be easily reached for any resonance model 
by simply splitting the correlator into a ``\,perturbative''\,  
and a ``\,non-perturbative''\, part (respectively $\Delta \Pi^{^{pert.}}$ 
and $\Delta \Pi^{^{non-p.}}$), and imposing the constraints in
Eq.~\eqn{eq.relationpQCD} and \eqn{eq.relationOPE}. 
     
What I would like to remark in this talk  is that the analysis must be taken one
more step forward;     
the correlators $\Pi^{^{N_C\to \infty}}$  carry  
extra  information   which can --and must-- be actually exploited.  
Deeper analyses may help to  discern between the different large $N_C$
resonance models present in the
bibliography~\cite{EspriuRegge,LVS,toymodel,Son,5D}.

In Ref.~\cite{correlator},   a set of sum-rules   
specially sensitive to the resonance parameters  was proposed:       
in order to compare the theoretical description to the experimental data 
we defined   the  moments 
\be
\mB^{^{(k)}}(Q^2) \,\,\,\, 
= \,\,\,\, (-1)^{k-1} \, \sqrt{\Frac{2k-1}{2}} \, 
\Int_0^\infty \, \Frac{2\, Q^2 \,\, dt}{(Q^2+t)^2} \,\, 
\,\,P_{k-1}\left[ \frac{t-Q^2}{t+Q^2}\right] \,\, \cdot  \,\, 
\Frac{1}{\pi} \mbox{Im}\Pi(t) \,\, ,  
\ee  
with $Q^2>0$, $k=1,2,...$ and where the  
$P_\ell[x]$ are the Legendre Polynomials ($P_0[x]=1,\, P_1[x]=x,...$).
As the order $k$ grows 
the Legendre Polynomials pinch the intermediate region  $t\sim Q^2$
in the integral    
whereas both the low and high  energy  extremes    
--where we rely on  the accurate experimental
data~\cite{Aleph} and QCD duality, respectively-- are enhanced.  
On the experimental side, we considered the $\tau$--decay 
spectral functions up to 
$t=3$~GeV$^2$ and pQCD beyond~\cite{correlator}. 
On the large~$N_C$ side,    
these sum-rules are absolutely convergent for any $k$,  
avoiding any problem of convergence in the resonance series.  
Moreover, when the pion   
pole is removed,   $\mbox{Im}\Pi(t)$ are bounded  functions, and   
then the moments  also result  bounded and obey the behaviour    
$\mB^{^{(k)}}(Q^2)\stackrel{k\to\infty}{\longrightarrow}0$~\cite{correlator}.

These moments are related to a combination of derivatives of the correlator in
the euclidean at
$t=-Q^2$ so, {\it a priori}, 
they can be computed through OPE~\cite{correlator}.  
However,  one can see in Fig.~\eqn{fig.B}  how  
the OPE is able to reproduce just the very first moments 
for  $Q^2=3$~GeV$^2$, breaking down
afterwards.  Indeed, it is quite a complicate task 
to separate     
the contributions from the anomalous dimensions and  
higher dimension condensates. It is no wonder  
the current controversy about the value  of high dimension
condensates~\cite{OPEcontro}.

By construction, the resonance descriptions reproduce pQCD and  
the first terms of the   OPE.  Furthermore, one 
can see in Fig.~\eqn{fig.B} that  
they are able to reproduce the experiment up to much higher moments.  
At this point, one must be aware of  the presence of    
next-to-leading order effects in $1/N_C$ due the  non-zero meson widths. 
Although the estimate of the subleading corrections is still under
investigation~\cite{incourse},   one finds  that 
the size of the required  corrections is much more  
important for the 5D--models  than for  the Regge--like  mass spectrum, what 
seems to favour the latter as the proper one.    
Nevertheless,  further  studies on alternative metrics   
that could solve this feature  of the current 5D--holographic models 
are really looked forward~\cite{Son}.

\section{Conclusions}
\tab
In this talk we have performed a matching of a large $N_C$ resonance
description to pQCD and the first terms of the OPE.  The matching 
to pQCD in
Eq.~\eqn{eq.relationpQCD} points out that 
imposing the asymptotic freedom
behaviour leads to a lack of control on the lightest
multiplet parameters;    the 
resonance series must be split into a ``\,perturbative''\,, 
$\Delta \Pi^{^{pert.}}$, 
and a ``\,non-perturbative''\, subs-series,  $\Delta \Pi^{^{non-p.}}$.

The second thing to remark is that , once performed the matching to pQCD and
OPE,  there is still extra information which admits being compared to
phenomenology.  Through the set of proposed sum-rules~\cite{correlator},  
one can see that the large $N_C$ resonance
theories are able to described a wider range of moments than OPE.  
This sort of studies 
can help to discern the proper structure of the hadronic mass
spectrum of QCD. Although the experimental uncertainties are of the percent
level, 
an estimated of the size of subleading corrections in $1/N_C$ is
crucial if one wants to perform a phenomenological analysis~\cite{incourse}.   

one needs still to estimate the size of the subleading corrections in
$1/N_C$~\cite{incourse}, in order to distinguish between resonance models.


\begin{thebibliography}{99}


 
\bibitem{Shifman}
M.A. Shifman {\it et al.}, {\it Nucl. Phys.} B {\bf 147} (1979) 385-447.    


\bibitem{NC}
G. 't Hooft, {\it Nucl. Phys.} B {\bf 72} (1974) 461;  
%
{\it Nucl. Phys.} B {\bf 75} (1974) 461;   
%
E. Witten, {\it Nucl. Phys.} B {\bf 160} (1979) 57.





\bibitem{correlator}
J.J. Sanz-Cillero, {\it  Nucl. Phys. } B {\bf 732} (2006) 136-168.



\bibitem{EspriuRegge}
S.S. Afonin  
{\it et al.},
{\it JHEP} {\bf  0404} (2004) 039.


\bibitem{LVS}
G.~Veneziano; {\it Nuovo Cimento} A {\bf 57} (1968) 190;
C.~Lovelace; {\it Phys. Lett.} B {\bf 28} (1068)  264; 
J.A.~Shapiro; {\it Phys. Rev.} {\bf 179} (1969) 1345. 
 
\bibitem{toymodel}
M. Golterman 
{\it et al.},  
{\it  JHEP} {\bf 0201} (2002) 024 ;
%
S.R. Beane, {\it Phys. Rev.} D {\bf 64} (2001) 116010 ;
%
M. Golterman and S. Peris, {\it JHEP} {\bf 01} (2001) 028 ;    
%
%
S. Peris  
{\it et al.}, 
{\it JHEP} {\bf 05} (1998) 011.
%
M.A. Shifman, 
hep-ph/0009131;   
%
O. Cat\`a {\it et al.},  
hep-ph/0506004. 




\bibitem{Son}
D.T. Son and  M.A. Stephanov;
{\it Phys. Rev.} D {\bf 69} (2004) 065020; 

\bibitem{5D}
L. Da Rold and  A. Pomarol; hep-ph/0501218;  
%
J. Hirn and V. Sanz, {\it JHEP} {\bf 0512} (2005) 030; 
%
J. Hirn 
{\it et al.}, 
hep-ph/0512240. 










\bibitem{PDG}
Particle Data Group (S. Eidelman et al.); 
{\it Phys. Lett.} B {\bf 592} (2004) 1.





\bibitem{OPEcontro}
V. Cirigliano
{\it et al.},   
{\it Phys. Rev.} D {\bf 68} (2003) 054013; 
%
S. Friot
{\it et al.},   
{\it JHEP}  {\bf 0410} (2004) 043; 
%
J. Bijnens 
{\it et al.},   
{\it  JHEP} {\bf 0110} (2001) 009;
%
J. Bordes {\it et al.}, 
hep-ph/0511293; 
%
S.~Narison, hep-ph/0412152.  





\bibitem{Aleph}
ALEPH Collaboration (S. Schael {\it et al.}),  
{\it Phys. Rept.} {\bf 421} (2005) 191-284 ;
%
CLEO Collaboration (S. Anderson {\it et al.}), 
{\it  Phys. Rev.} D {\bf 61} (2000) 112002 ;
%
OPAL Collaboration (G. Abbiendi {\it et al.}), 
{\it Eur. Phys. J. } C {\bf 35} (2004) 437-455;  
%
%
CMD-2 Collaboration (R.R. Akhmetshin {\it et al.}), 
{\it  Phys. Lett.} B {\bf 527} (2002) 161; 
{\it Phys. Lett.} B {\bf 578} (2004) 285;  
{\it Phys. Lett.} B {\bf 595} (2004) 101.   
%
BABAR Collaboration (B. Aubert {\it et al.}),  hep-ex/0502025;  


%
%


\bibitem{incourse}
J.J.~Sanz-Cillero, in preparation. 

\end{thebibliography}
\end{document}